# Enhanced MOKE magnetometry via Tunable Surface Plasmon resonance

Edgar J. Patiño[1*], Leidy Paola Quiroga S.[1] and César A. Herreño-Fierro[2]

[1] Departamento de Física, Superconductivity and Nanodevices Laboratory, Universidad de los Andes, Carrera 1 No. 18A-12, A.A. 4976-12340, Bogotá, Colombia

[2] Facultad de Ciencias y Educación, Universidad Distrital FJDC, Carrera 7 No. 40B - 53, 110231, Bogotá, Colombia

Here we demonstrate the enhancement of the transverse magneto-optical Kerr effect (TMOKE) signal, in the Otto configuration, due to surface plasmon resonance. We present an architecture where the low index dielectric has a variable thickness. This allows the plasmon resonance to be tuned, depending on the magnetic sample and thus be utilized to enhance the MOKE signal of samples separated from
the plasmonic device. We have achieved this by using air as low index dielectric where the evanescent wave extends, preceding to excitation of surface plasmons. The magnetic sample under consideration is a thin layer of cobalt coated by an ultrathin silver layer, on a silicon substrate (Ag/Co/Si). The sample is brought close enough to the prism//Air interface, allowing surface plasmon excitation in the air/Ag interface. This leads to an increase of the TMOKE signal up to ~ 2 per mil with respect to the incident light. This is about 8 times the traditional MOKE signal in the absence of plasmons. This is comparable with previous works using the Kretschmann-Raether configuration. This should find applications in MOKE-related technology by increasing its sensitivity to about one order of magnitude and allowing the magnetic sample to be separated from the plasmonic resonance device.

Keywords: Magnetometry, Transverse magneto-optical Kerr effect, Surface plasmon resonance.

## I. INTRODUCTION

Magnetometers are the key elements for the measurement of magnetic fields and magnetic dipole moments of magnetic materials. The measurement methods of the magnetic moment of a sample can be broadly classified into two categories: those which use magnetic induction produced by the sample stray fields, and those which optical properties depend upon intrinsic sample moment. For the first category currently, the broadly used techniques with the highest sensitivity are; the vibrating sample magnetometry (VSM) [1] and the superconducting quantum interference device (SQUID) [2]. These measure the magnetism of bulk samples and thus depend on sample volume. On the other hand, to directly examine the surface magnetic moment, the most commonly used technique is the magneto-optical Kerr effect (MOKE) [3], which uses light as a measuring probe. As oppose to the magnetic induction methods, the traditional MOKE setups include the possibility of imaging small magnetic regions at the micron and submicron scale. The sensitivity of this technique is limited to a very thin surface layer as a consequence of a narrow penetration depth of the incident light (few nanometres for most metals).

It has been widely demonstrated that the response of the MOKE signal from a magnetic layer can be enhanced as a result of giant electric fields produced by collective surface charge oscillations at a metal-dielectric interface [4] due to plasmon resonance. These oscillations are the result of the coupling of the parallel component of the wave vector between the external electric field and surface charges in the metal [5]–[7] leading to surface plasmon polariton (SPPs) waves. Until today most experimental work in this direction has been carried out exciting SPPs using a grating [8], [9] or in attenuated total reflection condition in so-called Kretschmann-Raether configuration [5]–[7]. These oscillations are achieved using a prism/metal/dielectric (P/M/D) layered structure where the prism has a high refractive index while the dielectric has a low refractive index.

In that configuration, the enhancement of the MOKE signal has been widely observed in structures where a thin ferromagnet (F) is located between two metals [10]–[12] i.e. (P/M/F/M/D). Here, the ferromagnetic layer provides the magneto-optic properties while the noble metal provides the plasmonic properties.

For practical applications as magnetometers, Kretschmann-Raether´s configuration is not suitable. This is because it requires the magnetic sample to be located between two metallic layers which makes it impossible to separate the sample from the measuring device. Furthermore, the thickness of the magnetic layer is limited to the total thickness of the M/F/M structure which cannot be more than the evanescent wave penetration depth.

[*] Corresponding author: epatino@uniandes.edu.co, Phone: +57 1 3394949, Fax: +57 1 3324516

In the present work, we demonstrate the solution by using the so-called Otto configuration where we use a P/D1/M/F/D2. A pending patent, related to this work, is currently in progress [13].

To our knowledge, there are only a few works in the Otto configuration that deal with magneto plasmonics. The first one was a theoretical proposition of a double layer dielectric and ferromagnetic metal [14] and experimental proof [15]. Later on, using gratings the Otto configuration was employed in references [8], [9], to characterize an ultra-thin magnetic film for applications in waveguides. In all these works; [8], [9], [14], [15], the sample was always part of the structure and the dielectric thickness was fixed in each structure.

For practical applications in magnetometry, a complete separation of the sample from the plasmonic structure is required. This is possible using the Otto configuration. However, in this scenario, the effective refractive index of the first dielectric D1 is modified with each sample. This complication can be overcome when the first dielectric thickness "$t$" is variable. Indeed, employing air as a low refractive index dielectric permits an easy variation of its thickness, and thus tune the effective refractive index of the structure. This way, although different magnetic samples bring variations in the plasmon resonance and TMOKE signal, the optimum resonance of the structure can be found by slightly modifying the angle of incidence and the dielectric thickness of the device.

## II. DEVICE ARCHITECTURE AND THEORY

As the first step to demonstrate these concepts we studied the effective structure prism/Air (~255 nm)//Ag (20 nm)/Co (10 nm)/Si (substrate) as depicted in Fig. 1.a. Here the double backslash indicates the prism/Air(~255 nm) is physically separated from the Ag (20 nm)/Co(10 nm)/Si (substrate) structure, i.e. in the Otto configuration. As we explain in section III the air gap of ~255 nm was the result of tuning the plasmon resonance, by adjusting the distance between prism and Ag surfaces, to obtain maximum TMOKE signal.

For convenience, the air gap between the prism and the Ag was achieved by fabricating a U shape frame of Ti (thickness ~ 500 nm), by optical lithography, on top of the Ag layer (Fig. 1b). However, it should be mentioned that it is also possible to grow the Ti frame/Ag (20 nm) structure directly on top of the prism leaving the Co(10 nm)/Si (substrate) completely separated from the plasmonic resonance device.

The metallic layers were grown on silicon substrate deposited by e-gun evaporation under high vacuum (UHV) conditions. The base pressure was better than $2 \times 10^{-6}$ Torr, and the evaporation pressure was less than $7\times 10^{-6}$ Torr. Each layer thickness was monitored during growth, by quartz balance, with a resolution better than 0.1 Å.

To study the angular dependence of the reflectance of the system, the prism was mounted onto a central rotating automated stage (goniometer) with a high resolution of better than 1 mrad, allowing the variation of the angle of incidence with a P polarized laser (532 nm).

To enhance the magneto-optical signal of the samples, separated a distance from the plasmonic device, we use the transverse magneto-optic Kerr effect (TMOKE).

In this configuration, the magnetic field is applied parallel to the plane of the structure and perpendicular to the plane of incidence. Details of the system can be found in reference [12].

When this configuration satisfies the total reflection condition at the Prism/Air interface, i.e., for angles greater than the critical angle, the evanescent field extends along the Air gap with exponential decay. The evanescent wave has an associated wave propagation constant parallel to the Prism/Air interface $K_{ev} = \frac{\omega}{c}\sqrt{\varepsilon_p}\sin\theta$. When evanescent field crosses the air gap and reaches the noble metal, at the resonance condition $K_{ev} = K_{SP}$, surface plasmons are excited at the Air/metal interface [16]–[18]. This produces electric fields of three or four orders of magnitude higher, that extend perpendicular to the metal surface and penetrates the thin noble metal layer to finally reach the magnetic Co sample, leading to the enhancement of the MOKE signal. Here we observe a strong signal enhancement close to twenty folds with respect to the sample signal in the absence of plasmonic excitation.

The relation between frequency and wave vector (dispersion relation) is unique for the system configuration. For example, for an Air/Metal semi-infinite interface the dispersion relation is given by the expression $K_{SP} = \frac{\omega}{c}\sqrt{\varepsilon_m\varepsilon_d/(\varepsilon_m + \varepsilon_d)}$ where $\varepsilon_m$ and $\varepsilon_d$ are the dielectric constants of the metal and dielectric medium respectively. For the case of magnetic samples instead of having a dielectric constant, we have a dielectric tensor that describes the MO activity. This can be characterized by different geometries responsible for the Magnetic Optic Kerr Effect (MOKE). For the present investigation, we have focused on the transversal configuration i.e., the magnetization perpendicular to the plane of incidence or TMOKE. Here the dielectric tensor of the ferromagnetic layer is described by;

$$\varepsilon_m = \begin{pmatrix} \varepsilon_{xx} & 0 & -g_y M_y \\ 0 & \varepsilon_{yy} & 0 \\ g_y M_y & 0 & \varepsilon_{zz} \end{pmatrix}. \quad (1)$$

This dielectric tensor depends on; magnetization ($M_y$), the magneto-optical $g_y$ factor (that depends on the material, the incident frequency and temperature). The diagonal terms ($\varepsilon_{xx}, \varepsilon_{yy}, \varepsilon_{zz}$) consider the merely optic response of the system.

Given that the dispersion relation $K_{SP}$ explicitly depends on the dielectric tensor $\varepsilon_m$, it is possible to obtain a connection between wave vector $K_{SP}$ and magnetization $M_y$ that permits control over the plasmonic properties of the structure employing an external magnetic field.

In magnetic transversal geometry, the magneto-optic signal consists of a change in the reflected intensity as a function of magnetization. In this way the T-MOKE signal is defined as;

$$\frac{\Delta R}{R} = \frac{R(+M) - R(-M)}{R(0)}, \quad (2)$$

where $R(\pm M)$ are the reflectivity in the positive and negative saturated magnetizations states respectively, and $R(0)$ the reflectivity in the absence of magnetization. However, we can define the T-MOKE signal as the change in reflectivity under magnetization inversion

$$\Delta R = R(+M) - R(-M). \quad (3)$$

This eliminates the artificial enhancement result of the minimum close to plasmon resonance [10]–[12].
For the theoretical analysis of the anisotropic layered structures, we used numerical calculations by the scattering matrix method [19] to compute the optical and magneto-optical response of the multilayer structures.
The refractive index of the prism/glass–substrate system was experimentally measured by a Brewster-angle experiment giving a value of 1.448, while the air refractive index was taken as 1.

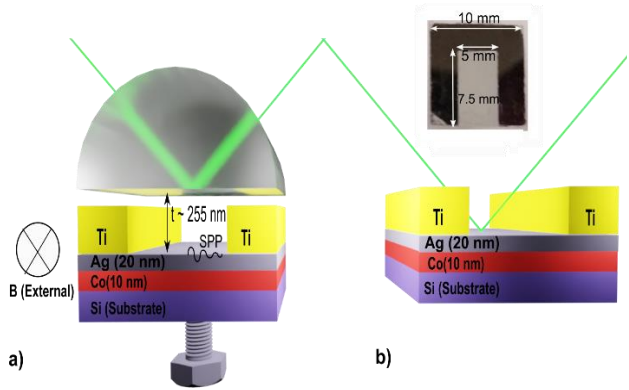

**Figure 1.** Enlarged schematic illustration of the studied structure, with prism and sample in a) the Otto configuration, (using a prism with a diameter of ~ 2.8 square cm) and b) the traditional MOKE configuration (light shining directly on the sample). The top view of 1 square cm used sample is shown in the inset, where U shape indicates Ti frame with dimensions.

### III. OPTICAL AND MAGNETO-OPTICAL CHARACTERIZATION

As shown in the schematic representation of the studied structure in Fig.1a, using a mechanical press the Ag(20 nm)/Co(10 nm)/Si (substrate) multilayer is put in contact with the prism, in the Otto configuration, supported on the Ti frame.
The pressure applied by the press slightly reduces the air gap between the Ag layer and the prism. This way is possible to tune the resonance condition until the minimum reflectivity value is found. For the chosen Ag/Co/Si(substrate) structure this optimum distance came to be about 255 nm as established from our data fit with the theory.

This structure was optically and magneto-optically (MO) characterized by angular spectral reflectivity and Transversal Magneto Optic Kerr Effect (T-MOKE) between incident angles of 33º and 65º. This was achieved by adopting a goniometer for the acquisition of the angular spectra reflectivity. The sample was illuminated through a cylindrical glass lens (BK7) by a p-polarized coherent light source of 532 nm wavelength (2.33 eV), while a photodetector collected data of the reflected light.

The magneto-optical characterization was done by an alternating magnetic field (B), with amplitude ~ 15 mT, located in the plane of the sample normal to the plane of incidence. In this way, the magnetization of the sample was along the sample plane. This allows extracting the T-MOKE signal from the changes in the reflectivity of the incident light due to the magnetic state of the sample. We used two methods for extracting small variations in the reflectivity signal; the first one using a Lock-in amplifier in phase with an oscillatory magnetic field. The second method simply uses amplifiers to obtain directly the hysteresis loops. A detailed explanation of these two experimental methods can be found in [12].

Fig. 2 shows the theoretical fit (dash line) and data (solid spheres) obtained from the SPR experiments in the angular reflectivity. Here the minimum of reflectance is found around 43º and the total internal reflection occurs around 35º.

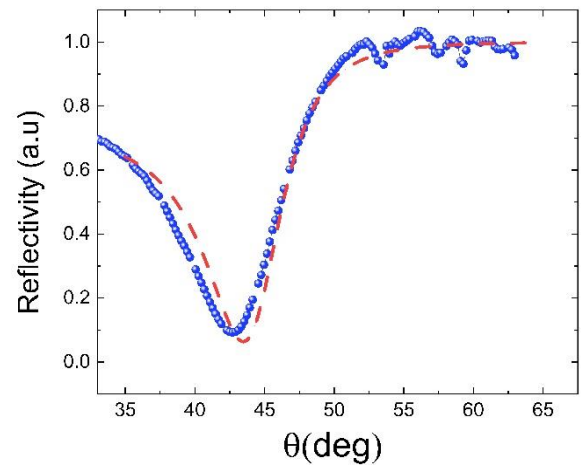

**Figure 2.** Angular reflectivity measurement (solid spheres) and theoretical prediction (dash line) for the sample in Otto configuration.

For the magneto-optical characterization, the first method uses a Lock-In amplifier where its voltage signal was referenced to an AC signal generator responsible for the external magnetic field. This way the system analyses only the intensity variations of light in phase with the frequency of the magnetic field. The experimental TMOKE signal obtained this way, can be found in Fig. 3 (solid spheres). The red dash line in this graph corresponds to the signal predicted by the theory. Furthermore, when the prism was completely removed as given in Fig. 1b and the light was illuminated directly on the sample, in a traditional MOKE configuration, the TMOKE signal was less than 0.1 per mil as shown in Fig. 3 using black dashed lines.

Here the maximum TMOKE signal shows an increase around 46º of ~2 per mil with respect to the incident light intensity. If we compare this value with the signal at 33º of ~ 0.3 per mil below the total reflection condition, in the absence of plasmon resonance this is about 7 times larger. This is comparable with previous works [10], [12], [20], different structures using the Kretschmann-Raether configuration.

measurements with the Lock-In amplifier (first method). In the presence of surface plasmons, we observed an augmented hysteresis loop (Fig. 4 solid spheres) in agreement with the enhancement observed with the TMOKE signal. Finally, a direct comparison of TMOKE hysteresis with the corresponding using traditional MOKE (Fig. 1b), can be found in Fig. 4. Here the light was directly irradiated towards the magnetic sample, giving a smaller hysteresis loop without the prism (Fig. 4 hollow spheres).

Given that in this case, the angle of incidence does not play any role on the MOKE signal we just took few measurements at the angle where the maximum TMOKE signal is observed ~ 46º. Fig. 4 shows the comparison of the hysteresis loops in the presence of SPR (solid spheres) and directly illuminated (hollow spheres) without a prism. Here the TMOKE signal shows an enhancement ~ 8 times larger, in the presence of surface plasmons, in comparison with the ones obtained with traditional MOKE.

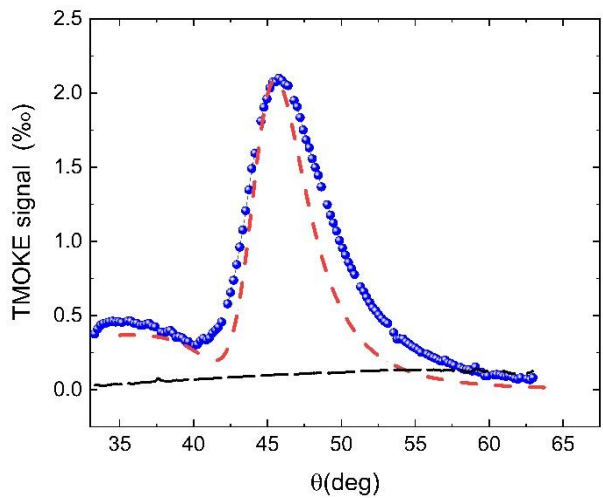

**Figure 3.** Transversal Magneto-Optical measurement (solid spheres) and theoretical prediction (red dash line) for the sample in Otto configuration. For comparison, the black dashed line shows the experimental data using the traditional MOKE configuration in Fig. 1b.

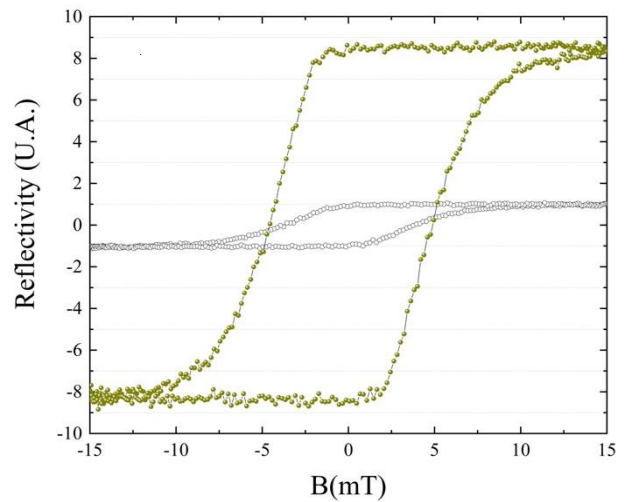

**Figure 4.** Hysteresis loops, taken at 46º, for the sample in presence of SPR (solid spheres) using the Otto configuration (Fig. 1a) and in the absence (hollow spheres) via direct illumination in traditional MOKE configuration (Fig. 2b).

## IV. RESONANCE VS TRADITIONAL MOKE CHARACTERIZATION

In order to directly compare the plasmon resonance MOKE enhancement with the actual hysteresis loops obtained with traditional MOKE, we used a second method of measurement. This allows us to directly observe the change of the hysteresis loops at each of the studied angles, accomplished by using a set of electronic amplifiers to obtain directly the hysteresis loops [12]. We should point out that each hysteresis loop was obtained at the same angle at which the TMOKE obtained signal data was obtained. This was taken right after the

## V. SUMMARY AND CONCLUSIONS

In this work, we have obtained enhancement of the TMOKE signal of the effective structure prism/Air (~255 nm)//Ag(20 nm)/Co(10 nm)/Si (substrate). This corresponds to the Otto configuration where surface plasmons have been exited in the Air/Ag interface across the air gap. Our results indicate an increase of the TMOKE signal of about ~ 2 per mil with respect to the incident light. This is about 8 times the MOKE signal in the absence of plasmons. This is comparable with our previous work [10], but this time in Otto configuration. Compared with conventional MOKE where light is directly irradiated towards the sample the TMOKE signal shows an enhancement ~ 8 times larger, in the presence of surface plasmons. The present work demonstrates the enhancement of the TMOKE signal by plasmon resonance of a magnetic sample that is physically separated from the plasmon resonance. The separation between the plasmon source and sample is achieved by an air gap. Given its principle of operation, does not depend upon sample magnetization direction. Therefore, this arrangement could be easily applied for other configurations too such as polar or longitudinal MOKE.

This development constitutes the practical implementation of a plasmon resonance magneto-optical Kerr effect (PRMOKE) in the transverse configuration. Here the Otto configuration allows magnetic samples to be separated by the plasmon resonance device, eliminating the necessity of optically coupling the samples to the prim.

One of the key advantages over traditional MOKE for applications is that light is not directly irradiated to the sample, with three consequences. The first one; laser light is not scattered by rough surfaces, therefore does not require a continuous and smooth sample. The second improvement is that it eliminates the production of local heat by the incident light. Finally, the third advantage, by controlling the distance between the plasmonic device and the sample, it is possible to tune the penetration depth of the plasmonic field into the sample. These characteristics make this development ideal to investigate different magnetic materials. This includes materials that cannot be deposited in transparent substrates.

This could find applications in magnetic sensors and magnetometers.

ACKNOWLEDGMENTS

We would like to acknowledge B. Garibello, who provided the MATLAB code for calculations, for initial code use assistance. This work was partially funded by Banco de la República No. 4.527, Convocatoria Programas 2012 y Fondo Publica y Expone of Vicerrectoría de Investigación y Creación, Facultad de Ciencias No. INV-2020-105-2036 - PROGRAMA 2021-2022 and "Convocatoria para la Financiación de Inversiones en Equipos de Laboratorio" Departamento de Física of Universidad de los Andes Bogotá, Colombia. C.H.F. thanks, CIDC Universidad Distrital, Miciencias (Colombia) and BMBF (Germany), project number 72874-2019, for partial funding.

DATA AVAILABILITY

The data that support the findings of this study are available from the corresponding author upon reasonable request.

[1] S. Foner, "Versatile and sensitive vibrating-sample magnetometer," *Rev. Sci. Instrum.*, vol. 30, no. 7, pp. 548–557, 1959, doi: 10.1063/1.1716679.

[2] R. C. Jaklevic, J. Lambe, A. H. Silver, and J. E. Mercereau, "Quantum interference effects in Josephson tunneling," *Phys. Rev. Lett.*, vol. 12, no. 7, pp. 159–160, Feb. 1964, doi: 10.1103/PhysRevLett.12.159.

[3] J. Kerr, "MOKE effect," *London, Edinburgh, Dublin Philos. Mag. J. Sci.*, vol. 3, no. 19, pp. 321–343, May 1877, doi: 10.1080/14786447708639245.

[4] Z. Q. Qiu and S. D. Bader, "Surface magneto-optic Kerr effect," 2000.

[5] E. Kretschmann, T. L. Ferrell, and J. C. Ashley, "Splitting of the dispersion relation of surface plasmons on a rough surface," *Phys. Rev. Lett.*, vol. 42, no. 19, pp. 1312–1314, 1979, doi: 10.1103/PhysRevLett.42.1312.

[6] S. A. Maier, *Plasmonics: Fundamentals and applications*. Springer US, 2007.

[7] H. Reather, "Surface plasmons on smooth and rough surfaces and on gratings," *Springer tracts Mod. Phys.*, vol. 111, pp. 1–3, 1988.

[8] O. V. Borovkova *et al.*, "TMOKE as efficient tool for the magneto-optic analysis of ultra-thin magnetic films," *Appl. Phys. Lett.*, vol. 112, no. 6, 2018, doi: 10.1063/1.5012873.

[9] O. V. Borovkova *et al.*, "Enhancement of the Magneto-Optical Response in Ultra-Thin Ferromagnetic Films and Its Registration Using the Transverse Magneto-Optical Kerr Effect," *Bull. Russ. Acad. Sci. Phys.*, vol. 83, no. 7, pp. 881–883, 2019, doi: 10.3103/S1062873819070098.

[10] C. A. Herreño-Fierro and E. J. Patiño, "Maximization of surface-enhanced transversal magneto-optic Kerr effect in Au/Co/Au thin films," *Phys. Status Solidi Basic Res.*, vol. 252, no. 2, pp. 316–322, Feb. 2015, doi: 10.1002/pssb.201451380.

[11] C. A. Herreño-Fierro, E. J. Patiño, G. Armelles, and A. Cebollada, "Surface sensitivity of optical and magneto-optical and ellipsometric properties in magnetoplasmonic nanodisks," *Appl. Phys. Lett.*, vol. 108, no. 2, Jan. 2016, doi: 10.1063/1.4939772.

[12] J. N. Hayek, C. A. Herreño-Fierro, and E. J. Patiño, "Enhancement of the transversal magnetic optic Kerr effect: Lock-in vs. hysteresis method," *Rev. Sci. Instrum.*, vol. 87, no. 10, Oct. 2016, doi: 10.1063/1.4966250.

[13] P. Q. E. J. Patiño, "Plasmonic device, system and method," 17/161195, 2021.

[14] T. Kaihara *et al.*, "Enhancement of magneto-

optical Kerr effect by surface plasmons in trilayer structure consisting of double-layer dielectrics and ferromagnetic metal," *Opt. Express*, vol. 23, no. 9, p. 11537, 2015, doi: 10.1364/oe.23.011537.

[15] T. Kaihara, H. Shimizu, A. Cebollada, and G. Armelles, "Magnetic field control and wavelength tunability of SPP excitations using Al2O3/SiO2/Fe structures," *Appl. Phys. Lett.*, vol. 109, no. 11, 2016, doi: 10.1063/1.4962653.

[16] J. R. Sambles, G. W. Bradbery, and F. Yang, "Optical excitation of surface plasmons: An introduction," *Contemp. Phys.*, vol. 32, no. 3, pp. 173–183, May 1991, doi: 10.1080/00107519108211048.

[17] A. K. Sharma, R. Jha, and B. D. Gupta, "Fiber-optic sensors based on surface plasmon resonance: A comprehensive review," *IEEE Sensors Journal*, vol. 7, no. 8. pp. 1118–1129, Aug. 2007, doi: 10.1109/JSEN.2007.897946.

[18] H. Ahn, H. Song, J. R. Choi, and K. Kim, "A localized surface plasmon resonance sensor using double-metal-complex nanostructures and a review of recent approaches," *Sensors (Switzerland)*, vol. 18, no. 1. MDPI AG, Jan. 01, 2018, doi: 10.3390/s18010098.

[19] B. Caballero, A. Garc\'\ia-Mart\'\in, and J. C. Cuevas, "Generalized scattering-matrix approach for magneto-optics in periodically patterned multilayer systems," *Phys. Rev. B*, vol. 85, no. 24, p. 245103, 2012.

[20] D. Martín-Becerra *et al.*, "Enhancement of the magnetic modulation of surface plasmon polaritons in Au/Co/Au films," *Appl. Phys. Lett.*, vol. 97, no. 18, Nov. 2010, doi: 10.1063/1.3512874.